\documentstyle[preprint,tighten,prd,aps,epsfig]{revtex}
\def\O{{\cal O}}
\def\Ot#1#2{\tilde{\cal O}^{#1}_{\bf #2}}
\def\Ht#1{{\tilde{\cal S}_{\bf #1}}}
\def\eval#1{\left\langle#1\right\rangle}
\begin{document}
\draft
\preprint{TAUP-2421-97}
\title{Ising description of the transition region in SU(3) gauge theory
at finite temperature}
\author{Benjamin Svetitsky and Nathan Weiss\thanks{On leave from the
Department of Physics and Astronomy, University of British Columbia, Vancouver, BC,
Canada V6T 1Z1.}}
\address{School of Physics and Astronomy, Raymond and Beverly Sackler
Faculty of Exact Sciences, Tel~Aviv~University, 69978~Tel~Aviv, Israel}
\date{4 May 1997}
\maketitle
\begin{abstract}
We attempt the numerical 
construction of an effective action in three dimensions
for Ising spins which represent the Wilson lines in the
four-dimensional SU(3) gauge theory at finite temperature.
For each configuration of the gauge theory, each spin is
determined by averaging the Wilson lines over a small neighborhood
and then projecting the average to $\pm1$ according to whether
the neighborhood is ordered or disordered.
The effective Ising action, determined via the lattice
Schwinger-Dyson equations, contains even (two-spin) and odd
(one- and three-spin) terms with short range.
We find that the truncation to Ising degrees of freedom produces
an effective action which is discontinuous across the gauge theory's
phase transition.
This discontinuity may disappear if the effective action is made
more elaborate.
\end{abstract}
\pacs{11.15.Ha 11.10.Wx 05.70.Fh}

\section{Introduction}
In studying the fluctuations of a complex statistical field theory,
it is frequently useful to define simple, effective degrees of freedom.
A wise choice of these degrees of freedom allows one to focus on specific
physics.
We are interested in the first-order phase transition of the SU(3) gauge
theory \cite{SU3,reviews}, specifically in surface phenomena such as
the surface tension between the phases \cite{surface,wetting} and the stability
of bubbles \cite{bubbles}.
Since the transition is first-order, all correlation lengths
stay finite on either side.
This means that there should be no delicate issues related to
renormalization-group fixed points and the symmetries that characterize
them.
We are thus led to degrees of freedom which simply specify whether
a given region is in the confined or unconfined phase.
Assigning values of $\pm1$ according to the two possibilities,
we are led to an effective theory of Ising spins.

This exercise is familiar from study of the liquid--gas 
transition \cite{Stanley}.
One assigns values to a local Ising spin $\sigma$
according to the local density $\rho$ of the fluid.
At the first-order phase boundary $T=T^*$ the density is discontinuous,
and this is reflected in a discontinuity in the magnetization 
$\langle\sigma\rangle$ of the effective Ising theory.
The simplest Ising action has only one term that is odd in the
spins, namely, the magnetic field term $h\sigma$.
The first-order transition occurs perforce at $h=0$, and
one identifies the liquid--gas phase boundary with the segment of
the $T$ axis between the origin and the Ising critical point
$T^{\text{Ising}}_{\text{cr}}$.
As one slides along the phase boundary towards the liquid--gas
critical point, the discontinuity in $\rho$ decreases until it
reaches zero at the critical point;
correspondingly, the discontinuity in $\langle\sigma\rangle$ vanishes
as one approaches $T^{\text{Ising}}_{\text{cr}}$.

The simplest Ising model, however, is incapable of describing an
arbitrary first-order phase transition.
An easy way to see this is to note that its $Z(2)$ symmetry implies
that
$\langle\sigma\rangle_{h\to 0^+}=-\langle\sigma\rangle_{h\to 0^-}$.
Since the correspondence between $\sigma$ and $\rho$ depends on
an arbitrary assignment in the first place, there is no reason to
suppose that this symmetry reflects accurately the values of $\rho$ 
on either side of $T^*$.
Hence there must be more than one odd term in the Ising action,
in order to move the transition away from $h=0$ and thus to destroy
the $Z(2)$ symmetry about the transition.

For the SU(3) confinement transition, we seek to characterize a neighborhood 
of a site~{\bf n} as confining or non-confining.
We naturally settle on the Wilson line~$L_{\bf n}$
as the quantity which does this.
As an order parameter of the $Z(3)$ symmetry of the Euclidean theory,
$L$ is discontinuous at the transition, with $\langle L\rangle=0$
in the confining phase below the transition and $\langle L\rangle\neq0$
in the plasma phase above.
We use the modulus of $L_{\bf n}$, suitably smeared, to assign a value
to $\sigma_{\bf n}=\pm1$.
(This smearing reflects the fact that confinement is a property
of a neighborhood, not of a point.)
Running a Monte Carlo simulation of the gauge theory, we generate
configurations of $L_{\bf n}$ which translate into configurations
of $\sigma_{\bf n}$.
We then use the Ising model's Schwinger-Dyson equations 
\cite{SDE,GA} to determine
an approximation to the effective Ising action $S_{\text{eff}}[\sigma]$.

This definition of $\sigma$ as a function of $L$ 
integrates over the $Z(3)$ dynamics of the Euclidean theory.
Since we are interested in understanding bubbles of the confining phase in the
plasma and vice-versa, distinguishing among the three ordered phases is
unnecessary.
Previous work \cite{Fukugita} has taken the opposite approach, projecting the 
complex Wilson line onto 3-state Potts spins $\tau=\exp 2\pi ni/3$.
This makes it harder to identify bubbles of the disordered phase
because they will only be visible when calculating averages of $\tau$
over sizable neighborhoods.
Since our $\sigma$ spins depend on the magnitude of $L$, they show directly
those places where $|L|$ is small and hence disordered.

Since $Z(3)$ domain structure plays a role in the confinement phase
transition, one might be concerned that domain walls should somehow
be represented in the effective action.
We note that calculations \cite{wetting}
have shown that, near the transition, the disordered phase wets the ordered
phases, and moreover that two ordered domains with different $Z(3)$
orientations will sandwich a disordered domain between them.
We note also that the physics of fluctuations among the three ordered phases
will influence the Ising couplings because the effective action comes
from an integration over {\em all} other degrees of freedom in the
gauge theory.
Nevertheless, integrating thus over the $Z(3)$ fluctuations 
may impair the ability of the Ising theory to
represent the phase transition correctly.
Our effective Ising action turns out to be discontinuous
at the phase transition.
Although this might be due to the small number of terms we permit in the action,
it might stem from the use of Ising variables itself.
We discuss this further below.

\section{Defining the effective Ising theory}

We simulate the SU(3) gauge theory at finite temperature
by using, as usual, a Euclidean lattice of $N_t\times N_s^3$ sites, with the
physical temperature given by $T=(N_ta)^{-1}$ in terms of the lattice
spacing $a$.
All results presented in this paper were obtained from lattices with
$N_t=2$.
The gauge theory is governed by the Wilson plaquette action,
\begin{equation}
S_W=\beta\sum_{n\atop{\mu<\nu}}\mbox{Tr}\, U_n^{\mu}
U_{n+\hat\mu}^{\nu}U_{n+\hat\nu}^{\mu\dag}
U_{n}^{\nu\dag}\ ,
\label{Wilson_action}
\end{equation}
with $U_{n}^{\mu}\in{\rm SU}(3)$.
The order parameter for the confinement phase transition is the
Wilson line, defined on a site {\bf n} of a three-dimensional lattice
via
\begin{equation}
L_{\bf n}=\mbox{Tr}\,\prod_{n_0=1}^{N_t}U_{({\bf n},n_0)}^0\ .
\label{Wilson_line}
\end{equation}

As discussed in the Introduction, we will use $L$ to define the
effective Ising spins $\sigma$ which will label a neighborhood
on the lattice as confining or non-confining.
A first attempt might be to define
\begin{equation}
\sigma_{\bf n}=\left\lbrace
\matrix{-1,&|L_{\bf n}|<r_\sigma\cr
1,&|L_{\bf n}|>r_\sigma\cr}
\right.
\label{sigma1}
\end{equation}
The problem with this is that $L_{\bf n}$ fluctuates violently
from site to site.
Even deep in the connfining phase, where $\langle L\rangle=0$,
$L_{\bf n}$ is by no means confined to a region around zero,
and in fact fills the entire wedge available to it in the complex
plane (see Fig.~\ref{fig1}).
$\sigma$ as defined by (\ref{sigma1}) thus does not offer a good
definition of a domain in the confining phase.

The fluctuations in $L_{\bf n}$ are reduced if it is averaged over
a small volume.
We define $L_{\bf n}^{[m^3]}$ to be the average of $L$ over the
$m\times m\times m$ block surrounding\footnote{If $m$ is even then
{\bf n} is a site of the dual lattice.}
{\bf n}.
A glance at Fig.~\ref{fig1} shows that $L_{\bf n}^{[8]}$ discriminates well,
on a local basis, between domains that resemble the two respective
phases.
$L_{\bf n}^{[27]}$, on the other hand, fluctuates too little about
the volume average of $L$, so that, with a reasonable value chosen
for $r_\sigma$, the $\sigma$ spins would lose all information
about fluctuations and remain entirely ordered with $\sigma=\pm1$.
We thus choose $L_{\bf n}^{[8]}$ for insertion in (\ref{sigma1})
to calculate the $\sigma$ configurations; as shown in Fig.~\ref{fig2},
we set $r_\sigma^2=0.8$.

With the definition of $\sigma$ in hand, we turn to the determination
of the effective Ising action.
In principle the action has an infinite number of terms;
we truncate it to a combination of a magnetic field term
and two- and three- spin terms with range 2,
\begin{equation}
S_{\text{eff}}[\sigma]=\sum_{\alpha}\beta_{\alpha}\O^{\alpha}\ ,
\label{Seff}
\end{equation}
where the seven operators $\O^{\alpha}$ are listed in Table~\ref{table1}.
The two-spin operators $\O^2$ and $\O^3$, as well as the three-spin
operator $\O^4$, couple spins within distance $\sqrt2$;
the remaining operators reach out to distance 2.

A Schwinger-Dyson equation of the Ising theory is
derived by flipping a spin $\sigma_{\bf n}$ in the
sum defining the expectation value of some operator.
For the operators in Table \ref{table1}, we have
\begin{equation}
\eval{\Ot{\alpha}n}=-\eval{\Ot{\alpha}n \exp{2\Ht n}}
\label{SDE}
\end{equation}
where we have defined $\Ot{\alpha}n$ to be those terms in
$\O^{\alpha}$ that contain $\sigma_{\bf n}$, and
\begin{equation}
\Ht n=\sum_{\alpha}\beta_{\alpha}\Ot{\alpha}n
\end{equation}
is the part of the action that contains $\sigma_{\bf n}$.
These are seven equations for determining the seven unknowns
$\beta_{\alpha}$.
After generating an ensemble of $\sigma$ configurations via
Monte Carlo simulation of the gauge theory,
we determine $\beta_{\alpha}$ iteratively as solutions of (\ref{SDE}).

As a consistency check, one may use the vacuum equation
\begin{equation}
1=\eval{\exp{2\Ht n}}
\label{vacuum}
\end{equation}
or the Schwinger-Dyson equation for any other operator in the
theory.
A more satisfying check, however, is to run a direct Monte Carlo
simulation of the Ising model with action (\ref{Seff}) to see if
the expectation values of $\O^{\alpha}$ as computed in the gauge
theory are reproduced.
This was the procedure we followed.
We calculated error estimates by subdividing the ensemble.

\section{Results and Discussion}

We simulated the SU(3) gauge theory on a lattice of volume
$2\times16^3$.
The confinement phase transition is in the neighborhood \cite{Kennedy}
of $\beta=5.09$, and we settled on the value $\beta=5.091$ after
seeing no tunneling between the coexisting phases in moderately
long runs at that coupling.

Straightforward application of the method described above gives an
Ising action for a three-dimensional lattice of volume $16^3$.
We show the couplings for this action, derived from ordered and
disordered runs at $\beta=5.091$, in Table~\ref{table2}.
In both cases, the action contains couplings with range
$\sqrt3$~and~2 ($\O^6$~and~$\O^7$, respectively) which are as strong
as the shorter-ranged two-spin couplings ($\O^2$~and~$\O^3$)
and compete with them in sign.\footnote{Note that the magnetic
field is $h=-\beta_1$, and that {\em negative} values for
$\beta_2$, $\beta_3$, $\beta_6$, and $\beta_7$ indicate
ferromagnetic couplings.}
This raises the suspicion that a longer-ranged Ising action is needed to
reproduce the Ising configurations correctly, and that the range-2
action is too crude a truncation.
This suspicion is confirmed by simulating directly the Ising model with
the couplings just derived.
As seen in Table~\ref{table3}, comparison of $\langle\O^\alpha\rangle$
with the averages from the gauge configurations shows poor 
agreement.\footnote{The violent disagreement for the ordered phase,
including even the sign of the magnetization,
suggests that the gauge theory's operator averages are to be sought in
a metastable phase of the Ising action.
We did not succeed, however, in reaching this phase with our Monte Carlo.
In the disordered phase, the positive magnetic field $h=-\beta_1$ prefers
a positive magnetization, but the positive three-spin couplings $\beta_4$
and $\beta_5$ compete with it and turn the magnetization negative.}

This problem was encountered by Deckert {\em et al.} \cite{Deckert} in
a calculation of the effective action for the $Z(2)$ gauge theory.
A solution, noted in \cite{GA}, is to perform a block-spin
transformation on the spins, so that the effective action
has twice the range.
We do this simply by decimating the Ising spins, already defined via smeared
averages, to an $8^3$ sublattice.
Solving the Schwinger-Dyson equations with the decimated configurations
gives the couplings shown in Table~\ref{table4}.
The longer-ranged two-spin couplings, $\beta_6$ and $\beta_7$, are
negligible, as is the straight three-spin coupling $\beta_5$.
Moreover, comparison of an Ising Monte Carlo simulation with the
gauge theory (see Table~\ref{table5}) now gives satisfactory agreement.

The effective Ising couplings shown in Table~\ref{table4} vary smoothly
as $\beta$ is varied on either side of the transition, but they are
discontinuous across the transition.
At $\beta=5.091$ we have, then, {\em two} actions, $S_{\text{cold}}$
and  $S_{\text{hot}}$, which are the limits of 
$S_{\text{eff}}[\sigma]$ from the ordered and disordered sides of the
transition.
Curiously, we find that $S_{\text{hot}}$ is at a point of phase coexistence,
that is, at a phase transition between phases with $\langle\sigma\rangle<0$
and $\langle\sigma\rangle>0$.
We show in Table~\ref{table5} the expectation values of the seven operators
$\O^\alpha$ for $S_{\text{cold}}$ and for both phases of 
$S_{\text{hot}}$.
For $S_{\text{hot}}$, the phase with $\langle\sigma\rangle<0$
describes well the expectation values
in the gauge theory on the disordered side of its transition.
The other phase of $S_{\text{hot}}$, of course, does not;
neither does it describe the ordered phase of the gauge theory.
The action $S_{\text{hot}}$ thus ``knows'' that it describes
a phase transition,
but it is capable of describing correctly only one of the phases.
$S_{\text{cold}}$ describes the ordered phase well, and shows
no phase coexistence.

The discontinuity in the effective action
is an example of the singularities that can result from
renormalization group transformations.
Griffiths and Pearce \cite{Griffiths}
noted that a blocked action might be
a singular function of the unblocked couplings even though the
blocking transformation is local.
Later work \cite{badRG} found discontinuities in the blocked action
associated with first-order phase transitions in the original
action.
It was conjectured that there may be different renormalization-group flows
resulting from the various metastable phases at a fixed coupling.
In view of theorems proven by van Enter, Fern\'andez, and Sokal \cite{Sokal},
however, such discontinuities are impossible in an effective action
which possesses finite range in the infinite-volume limit.
Our effective action, however, is approximate in that it contains a small
number of couplings.
Adding longer-ranged and multi-spin terms to the action will bring 
consistency with the theorem of van Enter {\em et al.} in one of two ways:
Either the couplings will become continuous \cite{GAS},
or the effective action(s) will acquire too many non-local terms in the
infinite-volume limit, meaning the statistical measure is non-Gibbsian.
In the first case, we will have an effective action well suited to describing
the phase transition;
in the second case, the conclusion will be that an Ising description of the
phase transition is impossible.
Deciding between these alternatives requires great numerical precision.

The discontinuity of the effective action is sensitive to the definition
of the effective degrees of freedom, just as singularities in the
renormalization group may be created or eliminated by different
choices of the block-spin transformation.
A more sophisticated definition of $\sigma_{\bf n}$, perhaps
using a Kadanoff kernel to associate it with the smeared $L_{\bf n}$,
may restore continuity, even without marked increase in the number of
interaction terms.
Note also that a reduction of the gauge theory to $Z(3)$ spins
in \cite{Fukugita} resulted in an action that is continuous
across the phase transition.
Perhaps a more complex effective spin, combining Ising with $Z(3)$,
will yield an effective action that offers both continuity and
a local description of confinement physics.

\section*{Acknowledgments}

We thank Doug Toussaint for giving us a copy of the MILC collaboration's
Monte Carlo code for the SU(3) lattice gauge theory,
a program employing overrelaxation with the Kennedy-Pendleton
heat bath algorithm.
We are indebted to Professor Kurt Binder for a very valuable
conversation, to Professor Jechiel Lichtenstadt for a discussion
of error analysis, and to the Weizmann Institute of Science for its continuing
hospitality.
This work was supported by the Israel Science Foundation under 
Grant No.~255/96-1.
Further support was provided by the Basic Research Fund of
Tel Aviv University.
N.~W. wishes to thank the National Sciences and Engineering Research
Council of Canada and the Israel Science Foundation
for their financial support.

%
%
\begin{figure}
\caption{Distribution of the Wilson line $L_{\bf n}$, averaged over 
$m\times m\times m$ cubes, in the complex plane.
Left: $\beta=5.0$ (disordered phase).
Right: $\beta=5.2$ (ordered phase).
Top to bottom: $m=1$ (single site), $m=2$, $m=3$.
The lattice size is $2\times8^3$.}
\label{fig1}
\end{figure}
\begin{figure}
\caption{$L_{\bf n}$ distributions for $m=2$, as in Fig.~\ref{fig1},
with circle $|L_{\bf n}|^2=r_\sigma^2=0.8$ superimposed.}
\label{fig2}
\end{figure}
%
%
\begin{table}
\caption{Operators $\O^\alpha$ appearing in the truncated effective
Ising action.}
\label{table1}
\begin{tabular}{ll}
single spin&
$\displaystyle \O^1=\vphantom{\sum_a^a}\sum_{\bf n}\sigma_{\bf n}$\\
nearest neighbor&
$\displaystyle \O^2=\vphantom{\sum_a^a}\sum_{\bf n}\sum_\mu \sigma_{\bf n}
\sigma_{{\bf n}+\hat\mu}$\\
next-nearest neighbor&
$\displaystyle \O^3=\vphantom{\sum_a^a}\sum_{\bf n}\sum_{\mu<\nu} \sigma_{\bf n}\sigma_{{\bf n}+\hat\mu
\pm\hat\nu}$\\
3-spin bent&
$\displaystyle \O^4=\vphantom{\sum_a^a}\sum_{\bf n}\sum_{\mu<\nu} \sigma_{\bf n}
\sigma_{{\bf n}\pm\hat\mu} \sigma_{{\bf n}\pm\hat\nu}$\\
3-spin straight&
$\displaystyle \O^5=\vphantom{\sum_a^a}\sum_{\bf n}\sum_\mu \sigma_{\bf n}\sigma_{{\bf n}-\hat\mu}
\sigma_{{\bf n}+\hat\mu}$\\
$3^{\rm rd}\hbox{ neighbor}$&
$\displaystyle \O^6=\vphantom{\sum_a^a}\sum_{\bf n}\sigma_{\bf n}
\sigma_{{\bf n}+\hat x \pm\hat y\pm\hat z}$\\
$4^{\rm th}\hbox{ neighbor}$&
$\displaystyle \O^7=\vphantom{\sum_a^a}\sum_{\bf n}\sum_\mu \sigma_{\bf n}
\sigma_{{\bf n}+2\hat\mu}$\\
\end{tabular}
\end{table}

\begin{table}
\caption{Couplings $\beta_\alpha$ for the (tentative) effective Ising action
on a $16^3$ lattice, for the ordered and disordered phases at
$\beta=5.091$.}
\label{table2}
\begin{tabular}{cdd}
$\alpha$&ordered&disordered\\
\hline
1&0.054(4)&-0.135(16)\\
2&-0.455(1)&-0.390(8)\\
3&-0.052(1)&-0.026(2)\\
4&-0.0056(6)&0.021(1)\\
5&0.0063(9)&0.022(2)\\
6&0.044(1)&0.045(4)\\
7&0.152(1)&0.132(3)
\end{tabular}
\end{table}
\begin{table}
\caption{Averages of $\O^\alpha$ in ordered and disordered phases 
of the gauge theory at
$\beta=5.091$, compared with results of Ising Monte Carlo 
for the couplings listed in Table~\ref{table2}.
Averages are normalized to 1.}
\label{table3}
\begin{tabular}{cdddd}
$\alpha$&\multicolumn{2}{c}{ordered phase}&
\multicolumn{2}{c}{disordered phase}\\
&gauge theory&Ising MC&gauge theory&Ising MC\\
\hline
1&0.185(7)&-0.849(2)&-0.877(1)&-0.696(6)\\
2&0.500(1)&0.812(2)&0.837(1)&0.708(4)\\
3&0.353(1)&0.775(2)&0.804(1)&0.626(5)\\
4&0.137(6)&-0.738(3)&-0.779(1)&-0.595(6)\\
5&0.141(6)&-0.742(2)&-0.781(1)&-0.602(6)\\
6&0.269(1)&0.756(3)&0.789(1)&0.583(6)\\
7&0.206(2)&0.743(3)&0.781(1)&0.552(7)
\end{tabular}
\end{table}
\begin{table}
\caption{Couplings $\beta_\alpha$ for the effective Ising action after
decimation to an $8^3$ lattice, for several gauge couplings surrounding
the phase transition.}
\label{table4}
\begin{tabular}{cddddd}
$\alpha$&
\multicolumn{1}{c}{$\beta=5.08$}&
\multicolumn{1}{c}{$\beta=5.091$}&
\multicolumn{1}{c}{$\beta=5.091$}&
\multicolumn{1}{c}{$\beta=5.1$}&
\multicolumn{1}{c}{$\beta=5.2$}\\
&&(disordered)&(ordered)&&\\
\hline
1&0.14(6)&0.02(5)&-0.027(2)&-0.054(2)&-0.26(3)\\
2&-0.21(2)&-0.24(2)&-0.132(1)&-0.131(1)&-0.12(1)\\
3&-0.043(3)&-0.050(3)&-0.023(1)&-0.021(1)&-0.025(3)\\
4&-0.011(2)&-0.015(2)&0.0043(4)&0.0028(4)&-0.0014(16)\\
5&0.003(5)&-0.001(7)&-0.003(2)&-0.0003(9)&-0.004(4)\\
6&-0.002(3)&-0.002(4)&-0.006(1)&-0.005(1)&-0.006(2)\\
7&-0.003(2)&0.003(7)&0.0016(7)&0.005(1)&0.004(3)
\end{tabular}
\end{table}
\begin{table}
\caption{Comparison of operator averages $\langle\O^\alpha\rangle$
in ordered and disordered phases of the gauge theory at $\beta=5.091$,
compared with the effective Ising
actions $S_{\text{cold}}$ and $S_{\text{hot}}$ simulated directly.}
\label{table5}
\begin{tabular}{cddddd}
$\alpha$&gauge theory&$S_{\text{cold}}$&
\multicolumn{2}{c}{$S_{\text{hot}}$}&gauge theory\\
&(ordered)&
&$\langle\sigma\rangle>0$&$\langle\sigma\rangle<0$
&(disordered)\\
\hline
1&0.188(7)&0.154(2)&0.9866(1)&-0.8799(1)&-0.878(1)\\
2&0.206(2)&0.203(1)&0.9739(1)&0.7849(2)&0.782(2)\\
3&0.123(2)&0.119(1)&0.9734(1)&0.7772(2)&0.774(2)\\
4&0.060(3)&0.049(1)&0.9615(1)&-0.7006(2)&-0.698(2)\\
5&0.063(3)&0.053(1)&0.9614(1)&-0.6994(2)&-0.697(2)\\
6&0.092(2)&0.086(1)&0.9733(1)&0.7751(1)&0.772(2)\\
7&0.080(2)&0.073(1)&0.9733(1)&0.7745(2)&0.772(2)
\end{tabular}
\end{table}
\newpage
\begin{figure}[t]
\centerline{\psfig{figure=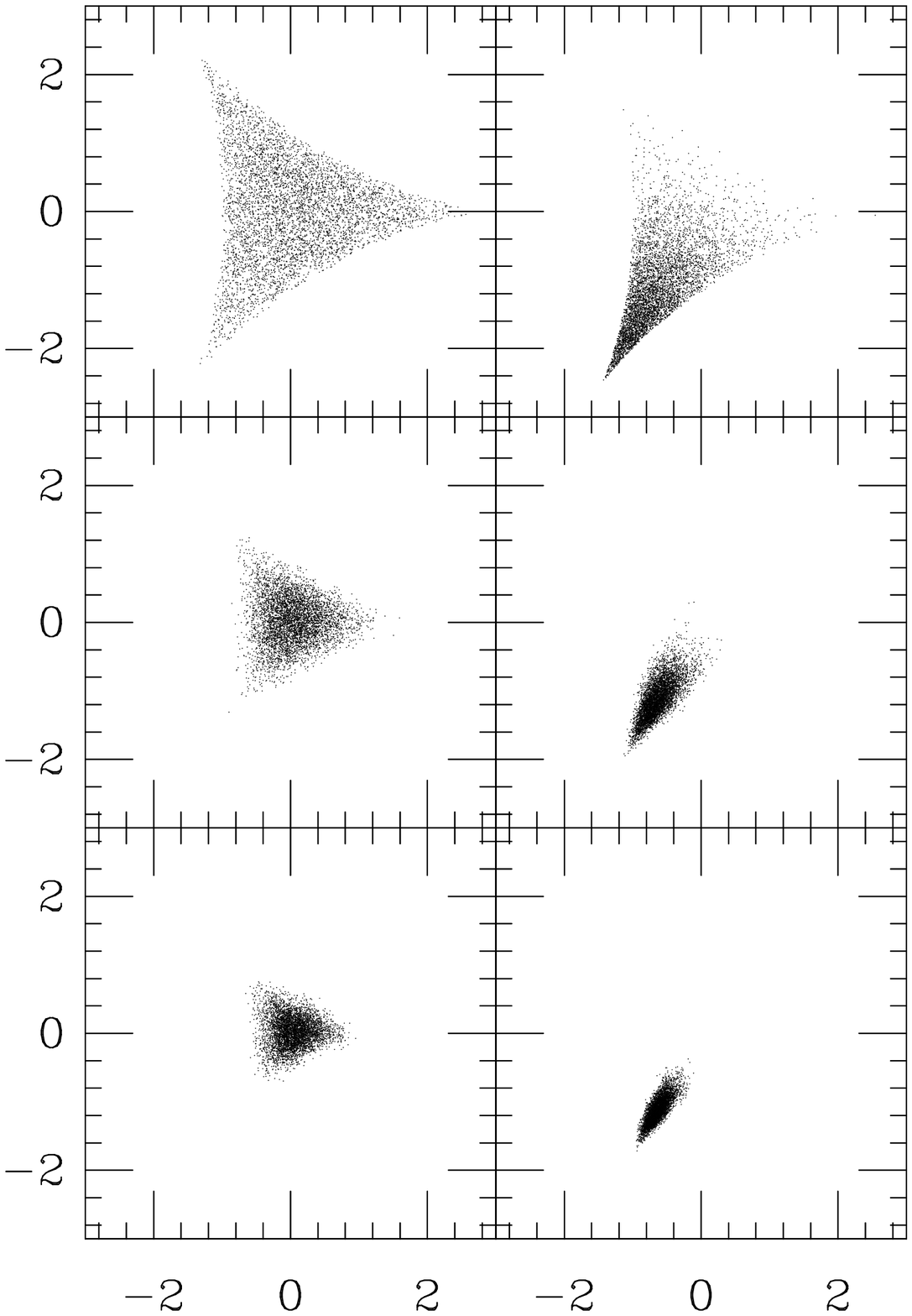,height=22cm}}
\centerline{FIG.~\ref{fig1}}
\end{figure}

\newpage
\begin{figure}[t]
\centerline{\psfig{figure=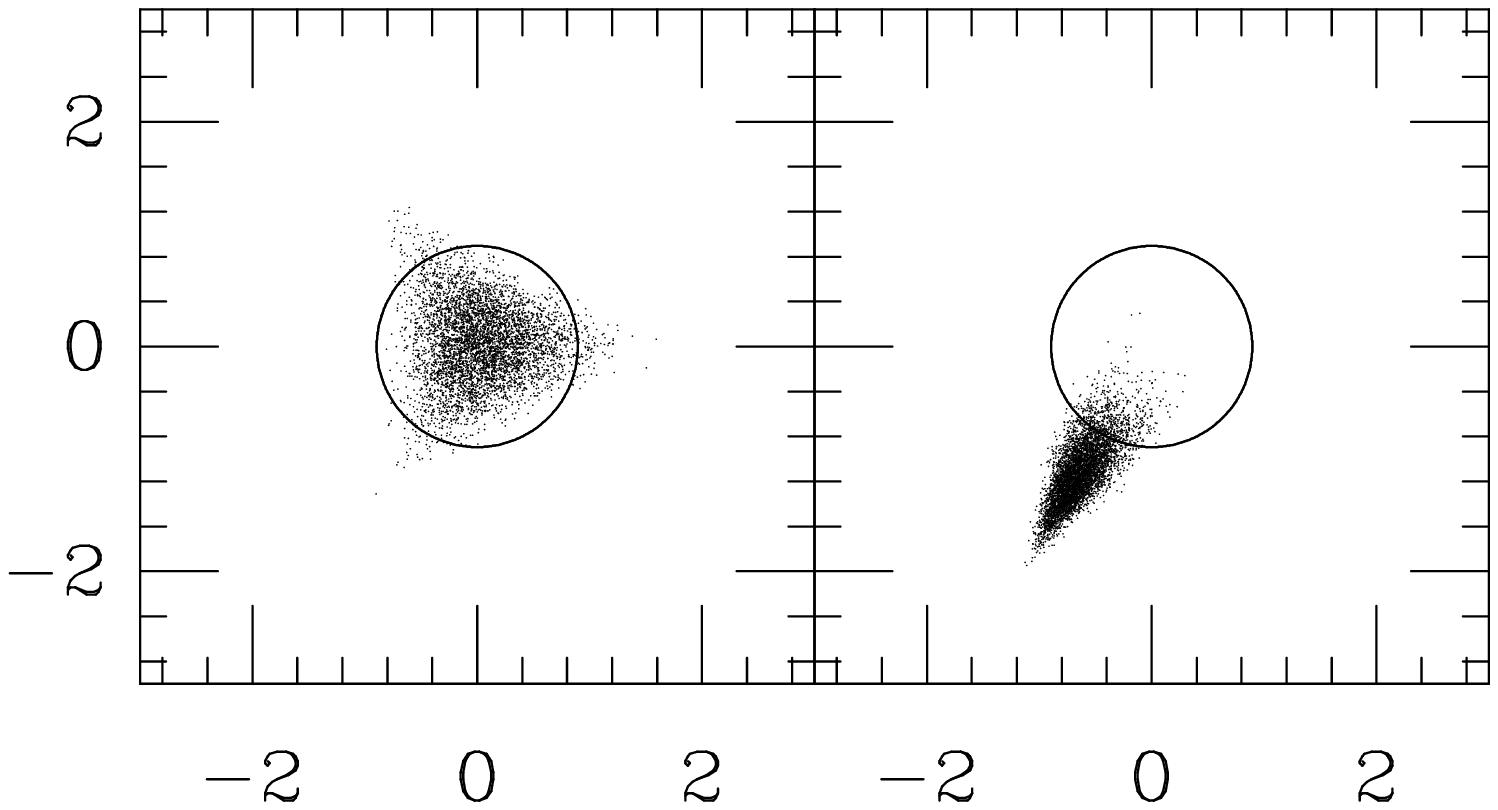,height=12cm}}
\centerline{FIG.~\ref{fig2}}
\end{figure}

\end{document}